# Pearls Are Self-Organized Natural Ratchets

Julyan H. E. Cartwright[1]

Antonio G. Checa[1,3]

Marthe Rousseau[2]

[1]Instituto Andaluz de Ciencias de la Tierra, CSIC–Universidad de Granada, Campus Fuentenueva, E-18071 Granada, Spain

[2]CNRS, UMR7561 Physiopathologie, Pharmacologie et Ingénierie Articulaires, Faculté de médecine, 9 Avenue de la forêt de Haye, BP 184, F-54505 Vandoeuvre les Nancy, France

[3]Departamento de Estratigrafía y Paleontología, Facultad de Ciencias, Universidad de Granada, E-18071 Granada, Spain

Version 3.6

**Abstract**

*Pearls, the most flawless and highly prized of them, are perhaps the most perfectly spherical macroscopic bodies in the biological world. How are they so round? Why are other pearls solids of revolution (off-round, drop, ringed), and yet others have no symmetry (baroque)? We find that with a spherical pearl the growth fronts of nacre are spirals and target patterns distributed across its surface, and this is true for a baroque pearl, too, but that in pearls with rotational symmetry spirals and target patterns are found only in the vicinity of the poles; elsewhere the growth fronts are arrayed in ratchet fashion around the equator. We demonstrate that pearl rotation is a self-organized phenomenon caused and sustained by physical forces from the growth fronts, and that rotating pearls are a — perhaps unique — example of a natural ratchet.*

**Introduction**

The only gemstones with a biomineral origin, since antiquity pearls have been prized for their beauty. Pearls are naturally formed by many species of molluscs in response to the presence of a foreign body such as a parasite. Nowadays they are deliberately cultured by introducing a nucleus into the organism, whereon biomineral is secreted around it. The majority of pearls, including those that are cultured for the jewellery industry using one of a few species of bivalve molluscs, are formed of nacre, or mother-of-pearl, one of the microstructures used by molluscs to construct their shells, a composite formed of some 95% calcium carbonate in its aragonite polymorph, together with 5% organic material: proteins, peptides, lipids and polysaccharides [1]. For the mollusc, it is the mechanical properties of nacre that are important: it is a strong material compared to its components [2], but to humans, it is the optical properties — the characteristic iridescence due to the structure of nacre — that cause pearls to be prized as jewels (Fig. 1).

Nacre consists of aragonite tablets with both extracrystalline and intracrystalline organic networks [3]. Shell nacre formation takes place in the extrapallial space that separates the mantle — the epithelial tissue of which secretes the compounds that compose nacre — from the shell (Fig. 2A). The extrapallial space is a thin liquid-filled gap, on the order of 100-200 nm wide, between the soft tissue of the organism and the uppermost interlamellar membrane that forms in this gap [1]. Interlamellar membranes of the polysaccharide chitin are laid down in this space in a process of liquid crystallization [4] and are then coated with protein before mineralization. Underneath the uppermost interlamellar membrane, individual tablets grow from a carbonate-charged silk phase through mineral bridges from the underlying layers [5] until they take the form of polygons some 300-600 nm thick and 5-10 μm in width [6,7,8]. Within tablets, nacre is a nano-composite: at the nanometre scale, the aragonite component inside tablets is embedded in a foam-like structure of organic materials in which the mean size of individual aragonite domains is around 50 nm [3]. In bivalves, on a pearl as on the shell, the different nacre layers grow at the same time producing a terraced pattern at the surface. This process leads, in bivalve shell nacre, to the surface having growth fronts arrayed in spiral and target patterns across it [9,4]; such patterns are a common feature of excitable media, of which nacre is an example [10].

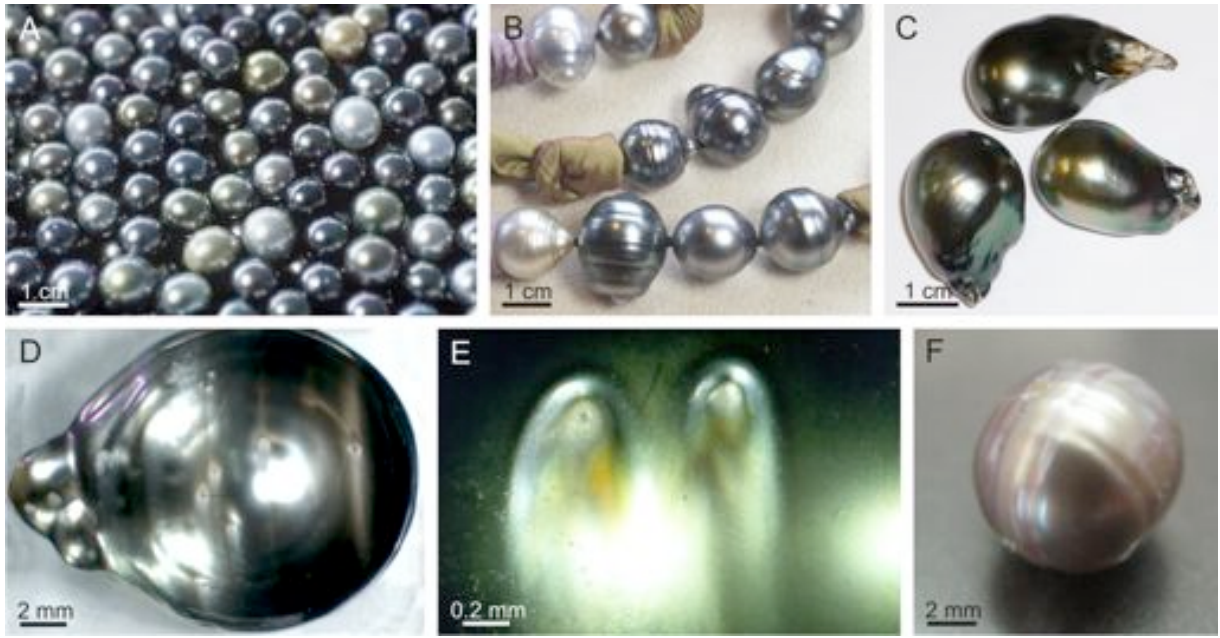

*Figure 1: Pearls and their macroscopic morphology: A, B, C; the three main categories of nacreous pearls. Cultured pearls of* Pinctada margaritifera*: A; high quality round pearls, B a necklace of drop and ringed pearls with rotational symmetry, C, baroque pearls without rotational symmetry. Rare examples of formation of sets of rings at angles; D,* Pinctada margaritifera *and F,* Pteria sterna*. E, Defects and accompanying wakes on the surface of a ringed* Pinctada margaritifera *pearl*.

The process of pearl formation, whether by natural means or through human intervention, is a response to an injury to the mantle tissue. In the case of natural pearls, mantle epithelial cells are displaced into connective tissue following damage to the mantle. They form a pearl sac (Fig. 2B) derived from the mantle epithelium, that secretes material forming a pearl in an analogous fashion to the shell itself. Pearl culturing techniques differ little between species; the general technique involves surgical implantation of a shell-derived nucleus, generally a sphere (Nu in Fig. 2), together with a section of mantle tissue removed from a donor of the same species. For *Pinctada* (saltwater pearls) the graft is into the gonad; for *Hyriopsis* (freshwater pearls) it is into the mantle. After grafting, this tissue develops to form a pearl sac. The secretion of the pearl sac, which is a single epithelium, is totally dependent on the physiology of the animal. Those cells programmed to secrete the components of the shell begin to deposit material on the nucleus. This secretion reproduces the structure of the shell in the pearl. The nacre, which is found in the interior of the shell, is thus on the contrary at the outside of the nacreous pearl. As long as the pearl remains in the animal, the pearl sac continues to secrete nacre-forming compounds and nacre lamellae progressively cover the nucleus, so the pearl grows (Fig. 2C, D), and the pearl sac expands with it.

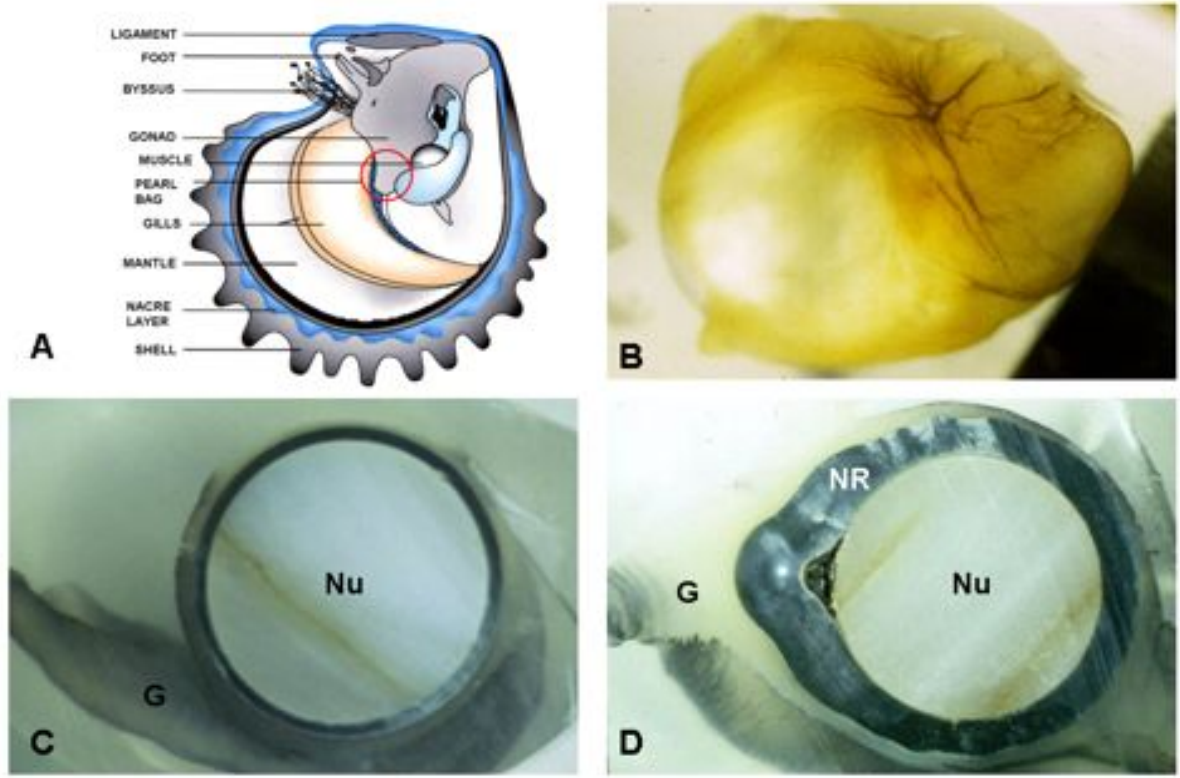

*Figure 2: The process of pearl formation: A The pearl sac and its position within the organism. B In toluene solution it is possible to see the pearl within the pearl sac. After sectioning the pearl inside the pearl sac we see the structure of the pearl inside for, C, a round pearl and, D, a pearl with rotational symmetry (Nu, nucleus ; NR, nacre ring ; G, gonad).*

**Results**

In a great majority of cases, cultured pearls can be placed, based on their overall form, into one of three categories: (1) spherical or round pearls, the most valued (Fig. 1A); (2) off-round, drop, and ringed pearls, having an axis of rotational symmetry, often showing rings about this axis (Fig. 1B); and (3) so-called baroque pearls, without any symmetry axis (Fig. 1C). We observe that, in an examination of the microscopic patterns of nacre growth on pearls (Fig. 3), those in categories (1) and (3) show many spiral and target patterns arrayed seemingly randomly over the surface, as in shell nacre. On the other hand, we find that those in category (2) display a quite different microscopic aspect, having spiral and target patterns in general only at the poles of the symmetry axis, while possessing, over the rest of the surface, approximately parallel growth fronts arrayed along lines of longitude (Figs 3A, B).

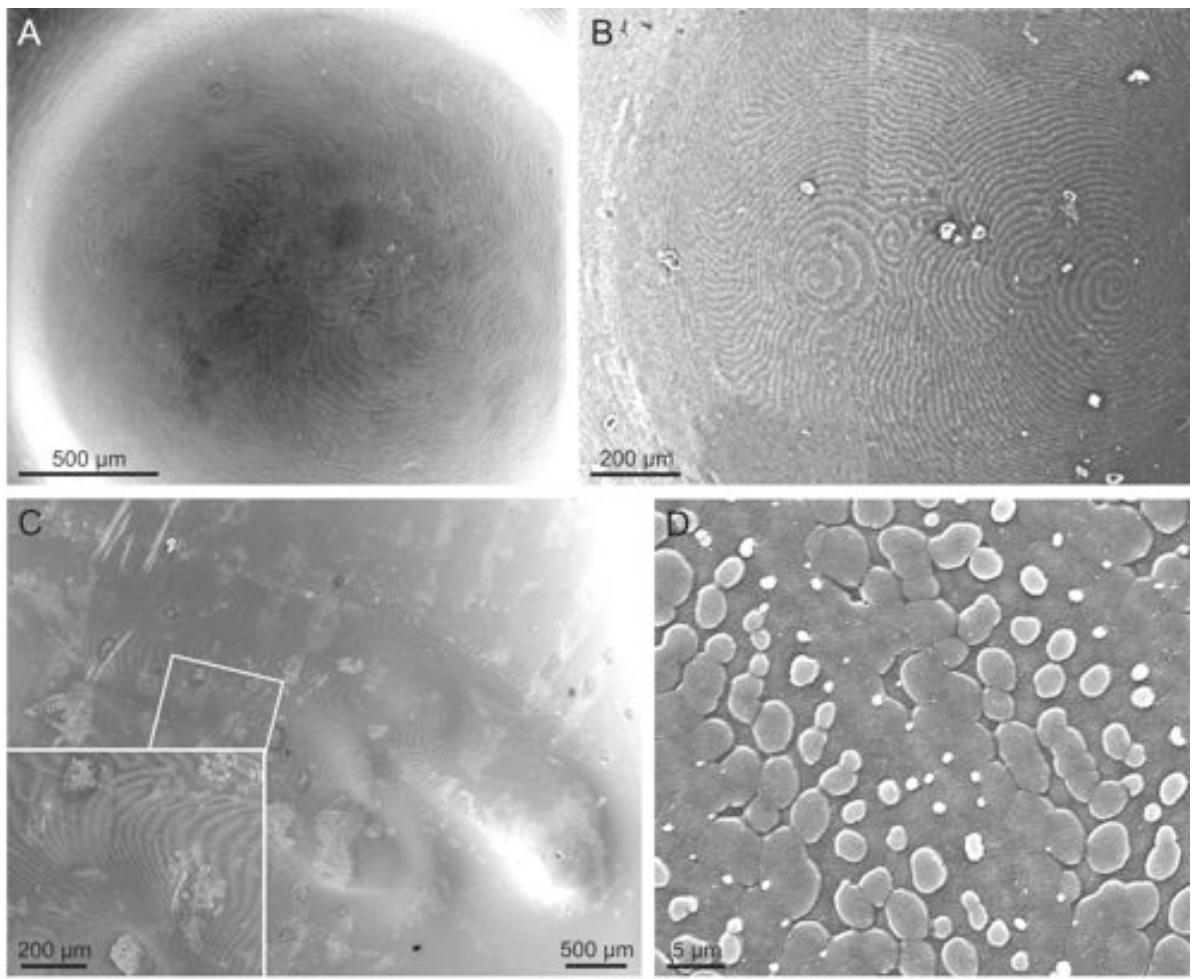

*Fig 3: Microscopic patterns of nacre growth: the arrangement of the growth fronts on pearls. Scanning electron micrographs of the surface of* Pteria sterna *pearls: A, B polar regions showing, A, a single central spiral and, B, several spirals. In both A and B there can be seen the interface between the polar and midlatitude regions with a transition to growth fronts that follow lines of longitude in ratchet fashion. C Pearl defect and its 'wake' showing growth fronts towards the lee. D Close-up view of growth fronts showing individual tablets.*

Cultured pearls, as we have pointed out, begin with a spherical nucleus, yet only a small percentage remain spherical to become round pearls. Those in categories (2) and (3) lose sphericity as they grow. Here we show that physics accounts for both the macroscopic morphology (the three categories of nacreous pearls) and the microscopic aspect (the arrangement of the growth fronts) of cultured pearls.

A crucial difference between shell and pearl nacre arises from the different geometries of the shell and the pearl: the pearl is immersed within its pearl sac and as such, unlike the shell, is free to move within it. It has long been suspected by pearl farmers that pearls turn within the sac [11,12]. This supposition has been confirmed through observation [13]; that unique study to date gave a rotation rate of once per 20 days: $10^{-6}$ Hz. In which sense is the rotation compared to the growth fronts? The question was not addressed by that study, but an answer may be given by examining depressional defects on the surface of otherwise rotationally symmetric pearls (Fig. 1E, 3C). These defects in the pearl surface appear not to be erosional features resulting from the removal of material within a pit, but, on the contrary, to be depressions caused by the deposition of material everywhere except in the pit, presumably owing to a local interruption in the supply of material from the epithelium; to a defect in the pearl sac. Such a defect would produce a diffusive depletion plume — that is to say, a plume arising from molecular diffusion, but one characterized by a reduced, rather than an elevated, concentration of solute compared to its surroundings — of material from the pearl sac across the extrapallial space towards the surface of the pearl. As a rotating pearl turns in the extrapallial liquid, the downstream wake of this depletion plume from molecular diffusion plus fluid flow will produce the morphology we observe of a steep slope as a leading edge and a shallower tail. From a close examination of the micrographs like that of Fig. 3C, we may establish that the growth fronts are aligned in the direction towards the tail of these defects; i.e., if the argument above is correct, they are are against the flow. We may note that a longer-lasting undersupply will produce the complete rings along lines of latitude around a rotationally symmetric pearl (Fig. 1B).

Is this correlation between flow direction and that of the growth fronts the cause, or the effect: Does rotation beget oriented growth fronts; or do oriented growth fronts beget rotation? We must ask what the forces are on a pearl. Firstly, what of exogenous forces? Considering in particular cultured saltwater pearls, as the gonad is not a muscular tissue, the pearl would not seem to be particularly susceptible to muscular contractions. While not denying that muscular forces exerted by the organism may affect the pearl, let us consider whether endogenous forcing could provide the motive force of its movement within the sac. The growth units of nacre, first chitin, then proteins, and later mineral, are added at the growth steps on its surface [1]. These growth units are chitin crystallites, protein molecules, and calcium carbonate, either in ionic form or else as aggregates; the chemical aspects are however not important in the following physical argument. Consider a growth unit impinging on a step from one side or the other: while there is a probability that it may stick when approaching from in front of a step, the probability of incorporation of the growth unit drops to almost nothing when a step is approached from the rear. We can see then that a step acts as a rectifier: it absorbs growth units from only one direction. However, the incorporation of growth units is only half the story; the majority of molecules incident on the step will be from the solvent, water, and they will carry away heat from the exothermic crystallization and solidification reactions at the growth front. The crucial point is that this heat will be carried off, as momentum, also in one direction (Fig. 4A).

The behaviour of the pearl then depends on the number and orientation of the steps on its surface. Consider a pearl that is growing from a spherical nucleus, which is already coated with nacre. The pressure exerted on the pearl's surface by the molecules as they rebound from the surface can be thought of as a species of osmotic pressure of the fluid on that surface (we say 'a species of' because osmotic pressure is usually considered for solute molecules impinging on a semipermeable membrane, while here both solute and solvent molecules impinge on the surface; nonetheless we find the physical analogy useful, and the mathematics is the same). If, instead of rebounding, a growth unit is adsorbed onto the surface, then the change in momentum of that growth unit is only $mv$, rather than $2mv$. Contrariwise, if, on rebounding, a molecule takes away more momentum than it arrived with, the change in momentum will be greater than $2mv$. This implies that the osmotic pressure is altered: in the absence of solvent, if all growth units that approach the surface are adsorbed, the osmotic pressure on that surface will be only half what it would be otherwise. However, that argument supposes a large 'sticking coefficient', and in fact the reverse is true: almost all admolecules or growth units do not stick and are not incorporated into the growing front but rebound from the surface (there seem to be few data on sticking coefficients, but for proteins, at least, they may indeed be very small [14]). More importantly still, we cannot ignore the solvent. If we take into account that the great majority of molecules rebounding from the surface will be from the solvent, and at the same time that the various processes of crystallization are exothermic, then instead of a lower osmotic pressure over a growth front, one will have a higher one: while only a small number of admolecules stick, molecules will on average rebound with higher energies after being heated by the surface.

In a pearl with many steps randomly oriented across its surface this increased osmotic pressure from growth will average out across the pearl, so it will not move, except for jiggling, or random rotations in the form of rotational Brownian motion, which will be negligible for a macroscopic pearl. But if the steps are oriented in general in a specific direction across the surface, then there will no longer be a cancellation of the momentum imparted to the pearl from averaging over the steps, and so the pearl may rotate. If, as we have discussed above, the osmotic pressure is increased in the direction towards the steps, the pearl will tend to rotate in the opposite direction, with the steps trailing (Figs 3C, 4A).

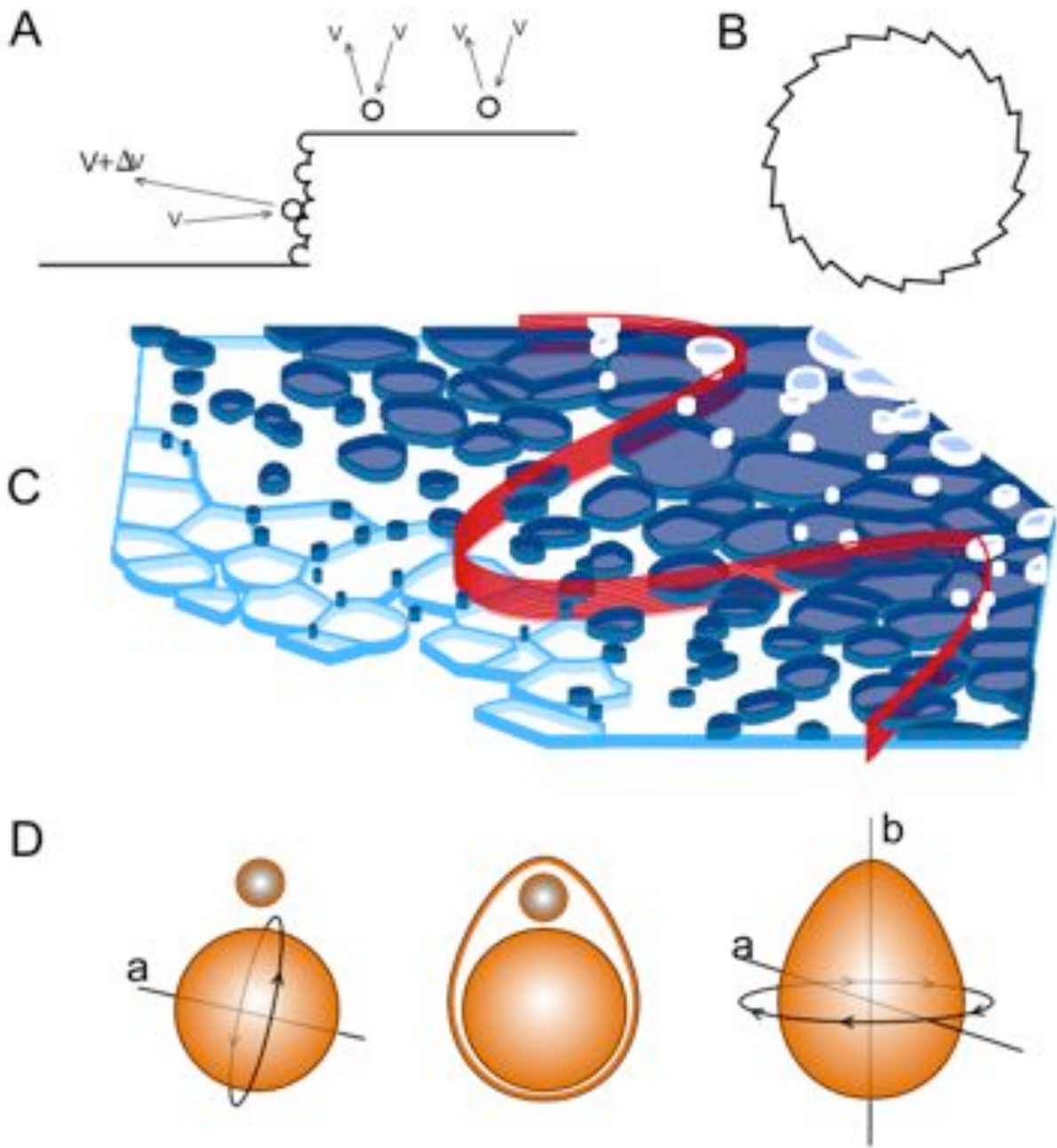

*Fig. 4 Mechanisms: A: Molecular forces originate from a growth step. B Steps form a ratchet around the circumference of a pearl. C Mullins-Sekerka dynamics leads to growth fronts oriented parallel to the flow. D Formation of pearl rings and sets of rings at angles.*

If we call the exothermicity coefficient from the various heats of solidification/crystallization, $X$, and the sticking probability, which depends on the propensity of molecules to stick, $p_s$, then the alteration in osmotic pressure ($PV=nRT$, $P$, pressure; $T$, temperature, $V$ volume, $n$, number of moles, $R = 8.3$ J K$^{-1}$ mol$^{-1}$, gas constant) is

$$P = ((1-p_s)X + p_s/2) \, nRT/V,$$

The area over which this pressure acts is the area of the vertical sections of the steps summed over the surface of the pearl. We can estimate this by considering the aspect ratio, $a$, of the steps; the height of the risers (vertical sections) $h$ compared to the length of the treads (horizontal sections) $l$. For nacre, typical values are $h \approx 0.5$ μm, $l \approx 20$ μm and $a = h/l = 0.025$. Across the surface of the sphere, the area of risers is then approximately $A \approx 4\pi r^2 a$, so the total force is

$$F = PA = 4\pi r^2 a \, ((1-p_s)X + p_s/2) \, nRT/V.$$

For example, for a 1 cm pearl ($r = 5 \cdot 10^{-3}$ m) at 300 K, growing from a solution with a molarity 50 M ($n/V = 50$ mol/L = 50 000 mol m$^{-3}$; pure water is approximately 55 M), with a sticking probability and exothermicity such that $(1-p_s)X + p_s/2 = 10^{-3}$ (we set these two parameters low to be conservative) $F = 4\pi(0.005)^2 \, 0.025 \, 10^{-3} \, 50\,000 \, 8 \, 300 \approx 0.1$ N. For comparison, the gravitational force on the pearl, $F = mg = 4/3\pi r^3 \rho g$, assuming its density about twice that of water, is $4/3\pi (0.005)^3 * 2000 * 10 = 0.01$ N: only one tenth of the growth force. As the growth force acts at a distance $r$ from the centre of the sphere, it causes a torque

$$\tau = PA\,r = F\,r,$$

which in our example is $\tau = 5 \cdot 10^{-4}$ N m. If we suppose the pearl to be a sphere rotating within another concentric sphere, the pearl sac, with radius $r = a+\varepsilon$, $a \gg \varepsilon$, it would rotate under this torque at a rate [15]

$$U = \tau(r^3-a^3)/(8\mu\pi r^3 a^3) \approx 3\tau\varepsilon/(8\mu\pi a).$$

If we assume the extrapallial liquid to be aqueous in terms of its viscosity, $\mu \approx 10^{-3}$ Pa s, and the extrapallial space $\varepsilon$ to be some 100 nm, we obtain $\cup \approx 3 \tau \varepsilon /( 8 \mu \pi a) \approx 3 \; 5.10^{-4} \; 10^{-7} / (8 \; 10^{-3} \pi \; 5.10^{-3}) \approx 10^{-6}$ Hz, which coincides with the unique observation of pearl rotation. Pearls certainly might move under muscular forcing too, but this result demonstrates that even without forcing from muscular contractions, which would presumably occur only at intervals, the physical forcing from growth will cause rotation. A few more clarifications ought to be made: one is that the geometry of the growth fronts in bivalve nacre is, in detail, complex, and involves both liquid as well as solid crystallization [1]; however it ought to be clear that the average effect will be the same as a simple step, so the above argument still holds. Secondly, in a rotating pearl, everything takes place in a rotating system, i.e., with fluid flow. However, the rotation velocity is so low that the Reynolds number is very small indeed, and the flow can be ignored for these purposes. Lastly, the rotation velocity is lower than the growth velocity of the steps, which is from one layer per day to one per hour, so the steps will advance as seen from the sac even as the pearl rotates in the opposite direction. However, this does not alter the physical argument.

The preceding argument has demonstrated that organized growth steps can cause rotation, but has so far failed to address one aspect of the question: How and why do the growth steps, in bivalve shell nacre organized into target and spiral patterns, become oriented in a nonrandom fashion across the surface in a rotating pearl? It is clear that around either a spiral or a target pattern, the growth steps are arrayed in an approximately circular manner. The effect of a spiral or target pattern is that the resultant force in this region of the surface will average to zero; to produce rotation, as few spirals/targets as possible are wanted on the pearl surface, with as much as possible of the surface covered with parallel (as far as that is possible on a sphere) steps following lines of longitude (i.e., arrayed from pole to pole). This is in fact what is observed in pearls: while spherical pearls are seen to have a large number of target and/or spiral patterns on their surfaces, nonspherical pearls with an axis of rotational symmetry are found to have the growth fronts aligned perpendicular to the growth direction *and* also in the same sense around the entire circumference, like the wheel of a ratchet, so that their effects add and do not cancel (Figs 3B, 4B). On these pearls we observe that spirals and target patterns are found at or close to the poles of the pearl (Fig. 3A); there must, at a minimum, be two such defects on the surface, the hairy ball theorem tells us (a hairy ball cannot be combed without leaving at least two whorls of hair sticking up [16]; equivalently a sphere covered in growth steps must have at least two defects in step direction). These observations imply that there is a dynamical mechanism present in this system such that, once rotation starts from some random perturbation, from a defect in the pearl sac to the intrusion of foreign material, there is then positive feedback so that the steps gradually align with the rotation axis. Such dynamics has been analysed for crystal growth steps in a flow [17]; that physics is equally valid here for liquid-crystal steps in nacre and is a form of the Mullins-Sekerka mechanism [18]. Higher concentrations of growth units upstream, and lower concentrations downstream, lead to the growth fronts stabilizing parallel to the flow (Fig 4C). This Mullins-Sekerka dynamics completes our physical argument by providing a self-organizing mechanism for the rotation. The rotation axis may change when a growing pearl encounters an adjacent small pearl, termed a keshi, that adheres to it, and this may alter the rotation axis to produce a rare pearl with two sets of rings at an angle to each other (Figs 4D, 1D, 1F). Lastly, a pearl that ceases to rotate owing to the presence of keshi or other defects may become a baroque pearl (Fig. 1C).

**Discussion**

The above argument for the motive force of a pearl is general for any crystal growth, so why do we not observe growing crystals to move under this force? For any conventional crystal lattice, the effect will cancel over the whole of the surface. The reason that it does not cancel for nacreous pearls is that nacre is formed from a liquid crystal with more flexibility than a solid crystal: the interlamellar membranes can be closer together and further apart, and this flexibility, not found in a solid crystal, is what enables growth fronts in a pearl to become organized perpendicular to the poles in ratchet fashion around the circumference. The physics we have described is also rather similar to in the Crookes radiometer. There, a macroscopic force, which Maxwell termed thermal creep, is induced from the edges of the hotter and colder sides of the vanes of a `windmill' in a rarefied gas [19,20,21,22]; perhaps Crookes' radiometer should be thought of as the first such ratchet in physics?

We are not aware of any other instance in which growth steps become aligned as a ratchet like that in a pearl, but the same physics that rectifies microscopic forces into macroscopic movement must operate in shell nacre, too, and must be important for transport of the components of nacre across the extrapallial space. Moreover, the possibility that pearls display, that by controlling, say, the temperature of a surface with a ratchet-like design one can produce movement of fluid or of an object itself is clearly of interest: that one could obtain a micromachine that could rotate in a given direction is a rather attractive idea that extends far beyond pearls to potential technological applications.

In summary, as soon as any perturbation to a spherical pearl initiates rotation, the growth fronts will begin to orient themselves parallel to the rotation axis via a form of Mullins-Sekerka instability and lock the pearl into rotation about the axis, which leads to the pearl becoming a non-spherical body of revolution, while if there are defects or other external factors that impede rotation, a pearl grows without any symmetry axis as a so-called baroque pearl. Hence the three main categories of cultured pearls: round pearls, pearls with rotational symmetry (off-round, drop, ringed), and baroque pearls. As well as explaining pearl morphologies, the understanding of the pearl as a natural ratchet should have interest for technological applications.

**Acknowledgements**

We thank Tahiti Perles and Douglas McLaurin for providing us with pearl samples and Stephan Engler for the pearl images of Fig 1A. We acknowledge funding provided by projects CGL2010-20748-CO2-01 and FIS2010-22322-C02-02 of the Spanish Ministerio de Ciencia e Innovación and the European COST Action TD0903.  All authors contributed in an integrated fashion to the work described in this paper; names are in alphabetical order.